\documentclass[10pt]{article}

\usepackage[margin=1in]{geometry}
\usepackage[T1]{fontenc}
\usepackage{lmodern}
\usepackage{microtype}
\usepackage{setspace}
\usepackage{lineno}

\usepackage{amsmath}
\usepackage{amssymb}
\usepackage{bm}

\usepackage{graphicx}
\usepackage{subcaption}
\usepackage{booktabs}
\usepackage{tabularx}
\usepackage{array}
\usepackage{float}
\usepackage{placeins}

\usepackage{algorithm}
\usepackage{algpseudocode}

\usepackage{titlesec}
\usepackage[numbers,sort&compress]{natbib}
\usepackage[
    colorlinks=true,
    citecolor=blue,
    linkcolor=blue,
    urlcolor=blue
]{hyperref}

\newcolumntype{C}[1]{>{\centering\arraybackslash}m{#1}}

\graphicspath{{pinn_paper_figures/}}

\titlespacing*{\section}{0pt}{6pt}{2pt}
\titlespacing*{\subsection}{0pt}{4pt}{1pt}
\titlespacing*{\subsubsection}{0pt}{3pt}{1pt}

\title{%
\vspace{-2cm}
{\large\textbf{Reconstruction of Magnetic Fields with a Physics-Informed Neural Network}}}

\author{%
\parbox{\textwidth}{%
\centering\normalsize
P. Adhikari$^{1}$, R. Bhattarai$^{1}$, J. Brewington$^{1}$, D. Buskirk$^{1}$, 
F. B. Hills$^{1}$, Md F. Hossain$^{1}$, R. McDonald$^{1}$,\\
P. A. Palamure$^{1}$, A. Reid$^{2}$, S. Sharmin$^{1}$,
A. Tewsley-Booth$^{1}$, and B. Plaster$^{1}$
}\\[0.6em]
\parbox{0.9\textwidth}{%
\centering
\small\textit{$^{1}$Department of Physics and Astronomy, University of Kentucky, Lexington, KY 40506, USA}\\
\small\textit{$^{2}$Department of Physics, Tennessee Tech University, Cookeville, TN 38505, USA}
}
}

\date{} 

\begin{document}

\maketitle

\renewcommand{\abstractname}{}
\begin{abstract}
\vspace{-\baselineskip}
\noindent\textbf{Accurate reconstruction of magnetic fields and their gradients is essential for assessing and controlling systematic effects in precision measurement experiments. We have developed a physics-informed neural network (PINN) method which reconstructs three-dimensional magnetic fields subject to the source-free Maxwell equations, $\nabla\cdot\mathbf{B}=0$ and
$\nabla\times\mathbf{B}=\mathbf{0}$. In an extension of our previous work on PINN methods for reconstruction of magnetic fields in inaccessible interior regions via measurements at exterior locations~\cite{Umit2022}, we now demonstrate the feasibility of applying PINN methods to the extraction of magnetic field gradient terms from discrete measurement points subject to constraints from Maxwell's equations.}
\end{abstract}


\section*{Introduction}

Accurate knowledge of magnetic field values and gradients is of paramount importance to controlling systematic errors in many precision measurement experiments, with one example being searches for a non-zero permanent neutron electric dipole moment (nEDM)~\cite{Chupp2019}. Recent state-of-the-art techniques for quantifying systematic errors in terms of general parametrizations of the magnetic field based on harmonic polynomial modes beyond the leading linear-gradient approximation~\cite{Pendlebury2004} have been presented in a series of foundational papers~\cite{Lamoreaux2005,Barabanov2006,Nouri2014,Brad2014, Nouri2015,Abel2019,Abel2022}.

In this paper, we extend our previous work~\cite{Umit2022} on magnetic field reconstruction using physics-informed neural networks (PINNs) to evaluate the potential for reconstruction of gradients. Our PINN model reconstructs the three-dimensional magnetic field,
\begin{equation}
    \mathbf{B}_{\mathrm{PINN}}(x,y,z)
    =
    \begin{pmatrix}
        B_x(x,y,z) \\
        B_y(x,y,z) \\
        B_z(x,y,z)
    \end{pmatrix},
\end{equation}
subject to the source-free Maxwell equations,
$\nabla\cdot\mathbf{B}=0$ and $\nabla\times\mathbf{B}=\mathbf{0}$.
We demonstrate the feasibility of extracting first order gradient terms relevant for systematic errors in nEDM experiments via PINN techniques through application to examples of magnetic field profiles developed for an nEDM experiment. As a benchmark of our PINN method, we compare the results from our PINN reconstruction with a spherical harmonic expansion.

\phantomsection
\section*{Physics-Informed Neural Network Method}
\label{sec:pinn_method}

\begin{figure}[htbp]
    \centering
    \includegraphics[width=0.99\linewidth]{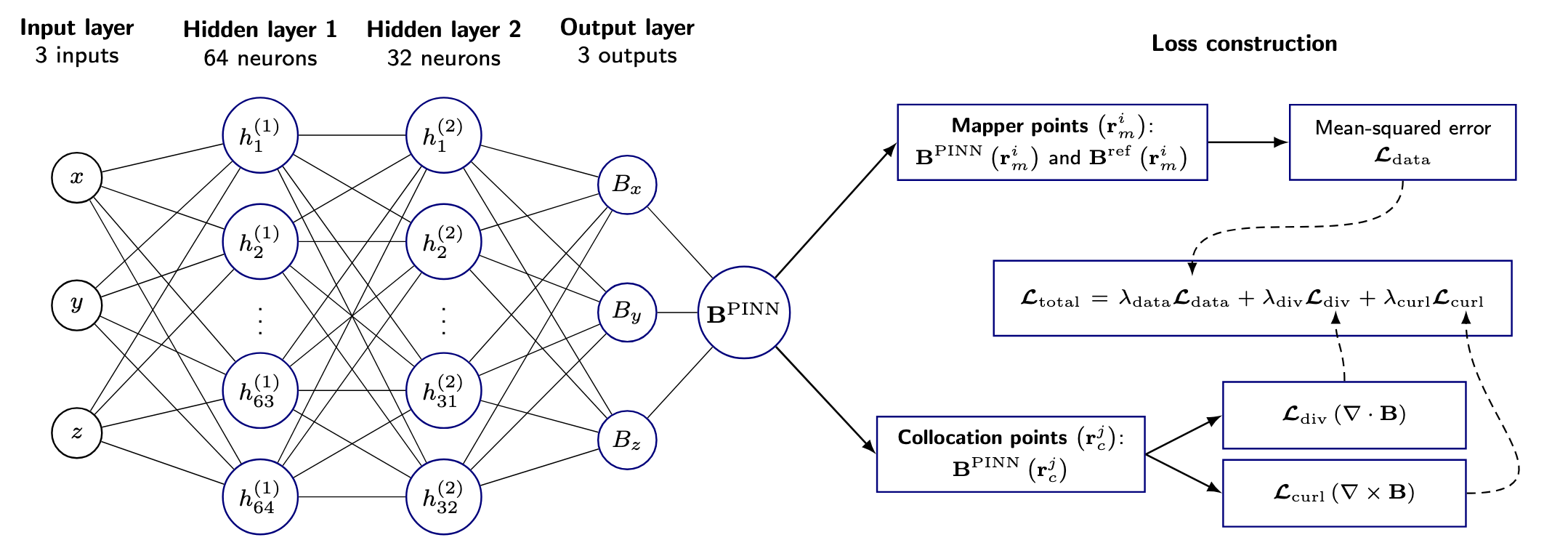}
    \caption{PINN architecture used for magnetic-field reconstruction. The network takes the spatial coordinates $(x,y,z)$ as inputs and predicts the magnetic-field components $(B_x,B_y,B_z)$, collectively denoted by $\mathbf{B}^{\mathrm{PINN}}$. The two hidden layers contain 64 and 32 neurons, respectively, with a hyperbolic tangent ($\tanh$) activation function. At the mapper locations, $\mathbf{B}^{\mathrm{PINN}}$ is compared with the known magnetic field, $\mathbf{B}^{\mathrm{ref}}$, to calculate the mean-squared-error data loss, $\mathcal{L}_{\mathrm{data}}$. The divergence and curl losses, $\mathcal{L}_{\mathrm{div}}$ and $\mathcal{L}_{\mathrm{curl}}$, are evaluated at the collocation points and combined with $\mathcal{L}_{\mathrm{data}}$ to construct the total loss, $\mathcal{L}_{\mathrm{total}}$.}
    \label{fig:network_structure}
\end{figure}

Physics-informed neural networks incorporate known physical laws directly into the training process by penalizing violations of those laws throughout the domain while minimizing the difference between the network predictions and the known data~\cite{Raissi2019}. For the reconstruction of a three-dimensional magnetic field in a source-free region, this involves enforcing $\nabla\cdot\mathbf{B}=0$ and $\nabla\times\mathbf{B}=\mathbf{0}$ through the definition of a loss function. The network used in this work takes the spatial coordinates $(x,y,z)$ as inputs and then outputs the corresponding magnetic field components $(B_x,B_y,B_z)$, as shown in Fig.~\ref{fig:network_structure}. 

In our work presented here, the spatial derivatives of the predicted field components with respect to the input coordinates were computed using automatic differentiation~\cite{baydin2018automatic_diff}, as implemented in PyTorch~\cite{pytorch2019}. These derivatives were then used to evaluate the divergence and curl of the reconstructed field throughout the domain.

The network was trained by minimizing a combined loss function consisting of a data-fitting term and physics-informed terms enforcing the divergence and curl constraints:

\begin{equation}
\mathcal{L}_{\mathrm{total}}
=
\lambda_{\mathrm{data}}\mathcal{L}_{\mathrm{data}}
+
\lambda_{\mathrm{div}}\mathcal{L}_{\mathrm{div}}
+
\lambda_{\mathrm{curl}}\mathcal{L}_{\mathrm{curl}},
\label{eq:total_pinn_loss}
\end{equation}

where $\lambda_{\mathrm{data}}$, $\lambda_{\mathrm{div}}$, and
$\lambda_{\mathrm{curl}}$ weight the contributions of the data, divergence, and curl loss terms, respectively, and can be adjusted based on the network performance.

The data loss was calculated as the mean-squared error between the PINN-predicted magnetic-field components and the corresponding known values. It was defined as
\begin{equation}
\mathcal{L}_{\mathrm{data}}
=
\frac{1}{3N_m}
\sum_{i=1}^{N_m}
\left[
\left(B_x^{\mathrm{PINN}}(\mathbf{r}_i)-B_x^{\mathrm{ref}}(\mathbf{r}_i)\right)^2
+
\left(B_y^{\mathrm{PINN}}(\mathbf{r}_i)-B_y^{\mathrm{ref}}(\mathbf{r}_i)\right)^2
+
\left(B_z^{\mathrm{PINN}}(\mathbf{r}_i)-B_z^{\mathrm{ref}}(\mathbf{r}_i)\right)^2
\right],
\label{eq:mse_loss}
\end{equation}
where $N_m$ is the number of measurement locations, $\mathbf{r}_i=(x_i,y_i,z_i)$ is the $i$th measurement location, and $\mathbf{B}^{\mathrm{ref}}(\mathbf{r}_i)$ denotes the simulated or measured reference magnetic field at that location.

 The divergence and curl losses were evaluated at a separate set of collocation points, $\mathbf{r}_c^j$, distributed throughout the reconstruction volume. These points allow the Maxwell constraints to be enforced throughout the domain rather than only where magnetic-field data are available. The divergence loss was defined as
\begin{equation}
\mathcal{L}_{\mathrm{div}}
=
\frac{1}{N_c}
\sum_{j=1}^{N_c}
\left[
\frac{\partial B_x}{\partial x}
+
\frac{\partial B_y}{\partial y}
+
\frac{\partial B_z}{\partial z}
\right]_{\mathbf{r}_c^j}^{2},
\label{eq:div_loss}
\end{equation}
and the curl loss as
\begin{equation}
\mathcal{L}_{\mathrm{curl}}
=
\frac{1}{N_c}
\sum_{j=1}^{N_c}
\left[
\left(
\frac{\partial B_z}{\partial y}
-
\frac{\partial B_y}{\partial z}
\right)_j^{2}
+
\left(
\frac{\partial B_x}{\partial z}
-
\frac{\partial B_z}{\partial x}
\right)_j^{2}
+
\left(
\frac{\partial B_y}{\partial x}
-
\frac{\partial B_x}{\partial y}
\right)_j^{2}
\right],
\label{eq:curl_loss}
\end{equation}

where $N_c$ is the number of collocation points and $\mathbf{r}_c^j$ denotes their positions. The numbers and spatial distributions of the measurement and collocation points within the reconstruction volume depend on the specific problem and are described under \hyperref[sec:specific_application]{Application to Specific Examples}.

\begin{algorithm}[H]
\caption{PINN training and magnetic-field reconstruction}
\label{alg:pinn_training}
\begin{algorithmic}

\Statex \textbf{Input:} Mapper data
$\{\mathbf{r}_m^i,\mathbf{B}_i^{\mathrm{ref}}\}_{i=1}^{N_m}$,
collocation points $\{\mathbf{r}_c^j\}_{j=1}^{N_c}$,
where $\mathbf{r}=(x,y,z)$,
$\mathbf{B}=(B_x,B_y,B_z)$, number of epochs $N_e$, and the data loss threshold $\varepsilon_{\mathrm{physics}}$

\Statex \textbf{Output:} Reconstructed field
$\mathbf{B}^{\mathrm{PINN}}(\mathbf{r})$
and its spatial derivatives

\For{$e=1,\ldots,N_e$}

    \State Evaluate
    $\mathbf{B}^{\mathrm{PINN}}(\mathbf{r}_m^i)$

    \State Compute $\mathcal{L}_{\mathrm{data}}$
    using Eq.~\eqref{eq:mse_loss}

    \If{$\mathcal{L}_{\mathrm{data}}
    \geq\varepsilon_{\mathrm{physics}}$}

        \State $\mathcal{L}_{\mathrm{total}}
        =
        \lambda_{\mathrm{data}}
        \mathcal{L}_{\mathrm{data}}$

    \Else

        \State Evaluate
        $\mathbf{B}^{\mathrm{PINN}}(\mathbf{r}_c^j)$

        \State Compute
        $\left.
        \dfrac{\partial\mathbf{B}^{\mathrm{PINN}}}
        {\partial\mathbf{r}}
        \right|_{\mathbf{r}_c^j}$
        using automatic differentiation

        \State Compute $\mathcal{L}_{\mathrm{div}}$ and
        $\mathcal{L}_{\mathrm{curl}}$ using
        Eqs.~\eqref{eq:div_loss} and~\eqref{eq:curl_loss}

        \State $\mathcal{L}_{\mathrm{total}}
        =
        \lambda_{\mathrm{data}}\mathcal{L}_{\mathrm{data}}
        +
        \lambda_{\mathrm{div}}\mathcal{L}_{\mathrm{div}}
        +
        \lambda_{\mathrm{curl}}\mathcal{L}_{\mathrm{curl}}$

    \EndIf

    \State Update the network parameters by backpropagating
    $\mathcal{L}_{\mathrm{total}}$

\EndFor

\State Reconstruct
$\mathbf{B}^{\mathrm{PINN}}(\mathbf{r})$
using the trained model

\State Compute
$\dfrac{\partial\mathbf{B}^{\mathrm{PINN}}}
{\partial\mathbf{r}}$
using automatic differentiation

\end{algorithmic}
\end{algorithm}

\subsection*{Spherical harmonic (SH) reconstruction}

To provide a benchmark for the results from our PINN methods, we also consider a spherical harmonic expansion, namely, that within the current-free measurement region, $\nabla\times\mathbf{B}=0$ implies $\mathbf{B}=-\nabla\Phi$, and together with $\nabla\cdot\mathbf{B}=0$ gives $\nabla^{2}\Phi=0$. The solution of Laplace's equation can then be expanded to represent the vector magnetic field components $B_x$, $B_y$, and $B_z$ at arbitrary three-dimensional locations. A similar technique for reconstructing the interior vector magnetic field from discrete measurements at fixed locations has been studied and presented for the nEDM experiment~\cite{Abel2019, Brad2014, Nouri2014, Nouri2015}.

Following the pioneering work of~\cite{Abel2019}, the magnetic field can be expanded as
\begin{equation}
\mathbf{B}(\mathbf{r})
=
\sum_{\ell=0}^{\ell_{\max}}
\sum_{m=-(\ell+1)}^{\ell+1}
G_{\ell,m}\,
\boldsymbol{\Pi}_{\ell,m}(\mathbf{r}),
\qquad
\mathbf{r}=(x,y,z),
\label{eq:field_expansion}
\end{equation}

where $G_{\ell,m}$ denotes the expansion coefficient of the corresponding vector harmonic-polynomial mode. The basis functions $\boldsymbol{\Pi}_{\ell,m}(\mathbf{r})$ are obtained from the magnetic potentials $\Sigma_{\ell+1,m}(\mathbf{r})$ as

\begin{equation}
\boldsymbol{\Pi}_{\ell,m}(\mathbf{r})
=
\nabla\Sigma_{\ell+1,m}(\mathbf{r}).
\label{eq:vector_basis}
\end{equation}

The magnetic potential is defined as

\begin{equation}
\Sigma_{\ell,m}(\mathbf{r})
=
C_{\ell,m}(\phi)\,
r^{\ell}P_{\ell}^{|m|}(\cos\theta),
\label{eq:scalar_basis}
\end{equation}
where $P_{\ell}^{|m|}$ is an associated Legendre polynomial. The explicit form and normalization of $C_{\ell,m}(\phi)$ are provided in~\cite{Abel2019}.

The components $B_x$, $B_y$, and $B_z$ were fitted simultaneously using modes up to $\ell_{\max}=3$, and although higher degrees of $\ell$ are, of course, possible, we found $\ell_{\max}=3$ was adequate  for the purposes of our work presented here. Because each magnetic-field mode of degree $\ell$ is derived from a scalar potential mode of degree $\ell+1$, the orders $\ell=0$, $1$, $2$, and $3$ contain $3$, $5$, $7$, and $9$ modes, respectively, resulting in 24 fitted expansion coefficients.

The SH method was used as a benchmark for evaluating the PINN because it provides an analytical representation of the magnetic field that satisfies the source-free Maxwell equations within the current-free measurement region. The expansion coefficients were determined by simultaneously fitting the measured $B_x$, $B_y$, and $B_z$ components. Once fitted, the SH expansion was used to reconstruct the magnetic-field components, and the corresponding spatial derivatives were calculated analytically using the fitted expansion coefficients. This enabled a direct comparison of the three reconstructed magnetic field components, $B_x$, $B_y$, and $B_z$, and the nine first order spatial gradients obtained using the PINN and  SH  methods.

\section*{Application to Specific Examples}
\label{sec:specific_application}

The magnetic field datasets considered in this work consist of a calculated map from a finite-element analysis simulation of a highly-uniform magnetic field designed for an nEDM experiment~\cite{Ryan2018,Jared2023,Amara2023} and measured data from bench-tests of a set of spin-transport coils (STC) designed for adiabatic spin transport~\cite{Crawford2021,Crawford2021_design, STC_inpreparation} of ultracold neutrons in this same nEDM experiment. We employed these two specific examples as the magnetic field profiles are very different: the former is, as indicated, highly uniform whereas the latter is highly non-uniform. Thus, these two examples demonstrate the utility of our technique for a wide range of field uniformities.

\subsubsection*{Simulated Highly-Uniform Magnetic Field}

COMSOL Multiphysics\textsuperscript{\textregistered}~\cite{COMSOL61} was used to simulate the highly-uniform magnetic field. The as-designed coil is a multigap solenoid consisting of eight octagonal sections designed to produce a highly uniform magnetic field of approximately $1~\mu\mathrm{T}$, as reported previously~\cite{Ryan2018,Jared2023,Amara2023}. The simulation provided the field components $(B_x,B_y,B_z)$ on a Cartesian grid $(x,y,z)$. In addition to simulations of the nominal baseline field, additional maps with Maxwell-consistent perturbative fields were also added to the COMSOL-simulated field for evaluating the field-reconstruction and gradient-extraction capabilities of the PINN under perturbative conditions, which were taken to be asymmetric.

\begin{figure}[htbp]
    \centering
    \includegraphics[width=0.9\linewidth]
    {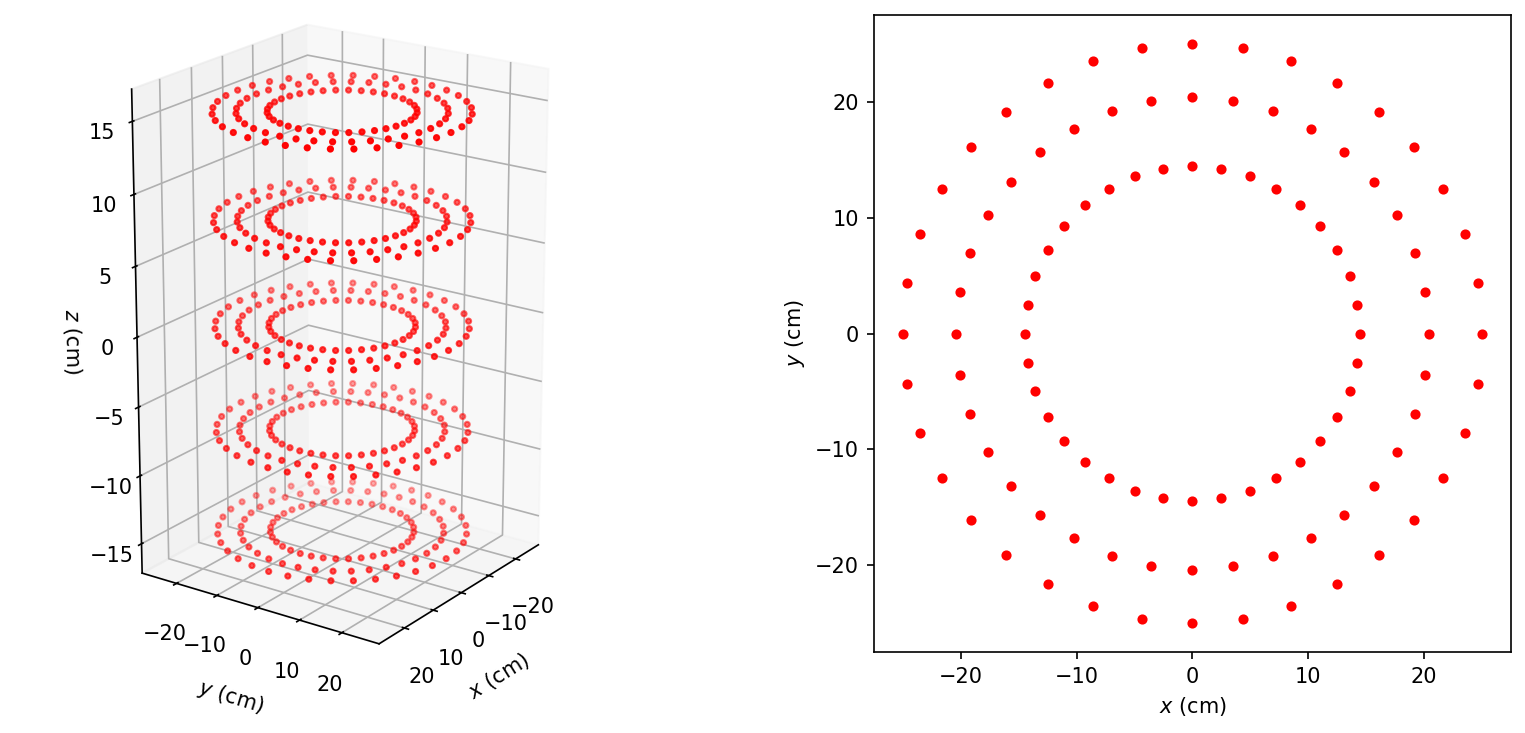}
    \caption{Magnetic-field sampling geometry for the simulated highly-uniform field. The left panel shows the 540 sampling locations within the cylindrical reconstruction volume, and the right panel shows their distribution in the $z=0$ plane.}
    \label{fig:measurement_geometry}
\end{figure}

For our work reported here, the area of interest within this highly-uniform magnetic field region was a cylindrical volume of radius $R = 25~\mathrm{cm}$ and axial extent $z = \pm 15~\mathrm{cm}$. Then, within this region of interest, and following the pioneering work of~\cite{Abel2022}, we defined a sampling geometry, shown in Fig.~\ref{fig:measurement_geometry}, which consisted of three radial rings, five axial planes, and 36 uniformly spaced azimuthal positions on each ring. The ring radii were defined as \(r_i=R\sqrt{i/3}\) for \(i=1,2,3\), dividing the circular mapping plane into equal-area radial regions. The same number of azimuthal points was assigned to each equal-area region to provide approximately uniform sampling across the mapping plane. The five axial planes sampled the central, intermediate, and end regions of the reconstruction volume.  This arrangement resulted in $5\times3\times36=540$ sampling locations, as shown in Fig.~\ref{fig:measurement_geometry}. The perturbed COMSOL-simulated field was interpolated at these locations to obtain the corresponding components $(B_x^{\mathrm{sim}},B_y^{\mathrm{sim}},B_z^{\mathrm{sim}})$. The resulting coordinates and simulated field values, denoted by $\mathbf{B}^{\mathrm{sim}}$, formed the reference dataset used for subsequent magnetic-field reconstruction and gradient extraction. The data loss was calculated as the mean-squared error between the PINN-predicted field, $\mathbf{B}^{\mathrm{PINN}}$, and the simulated magnetic-field components, $\mathbf{B}^{\mathrm{sim}}$, at the mapper locations.

\begin{figure}[htbp]
    \centering
    \includegraphics[width=0.95\linewidth]
    {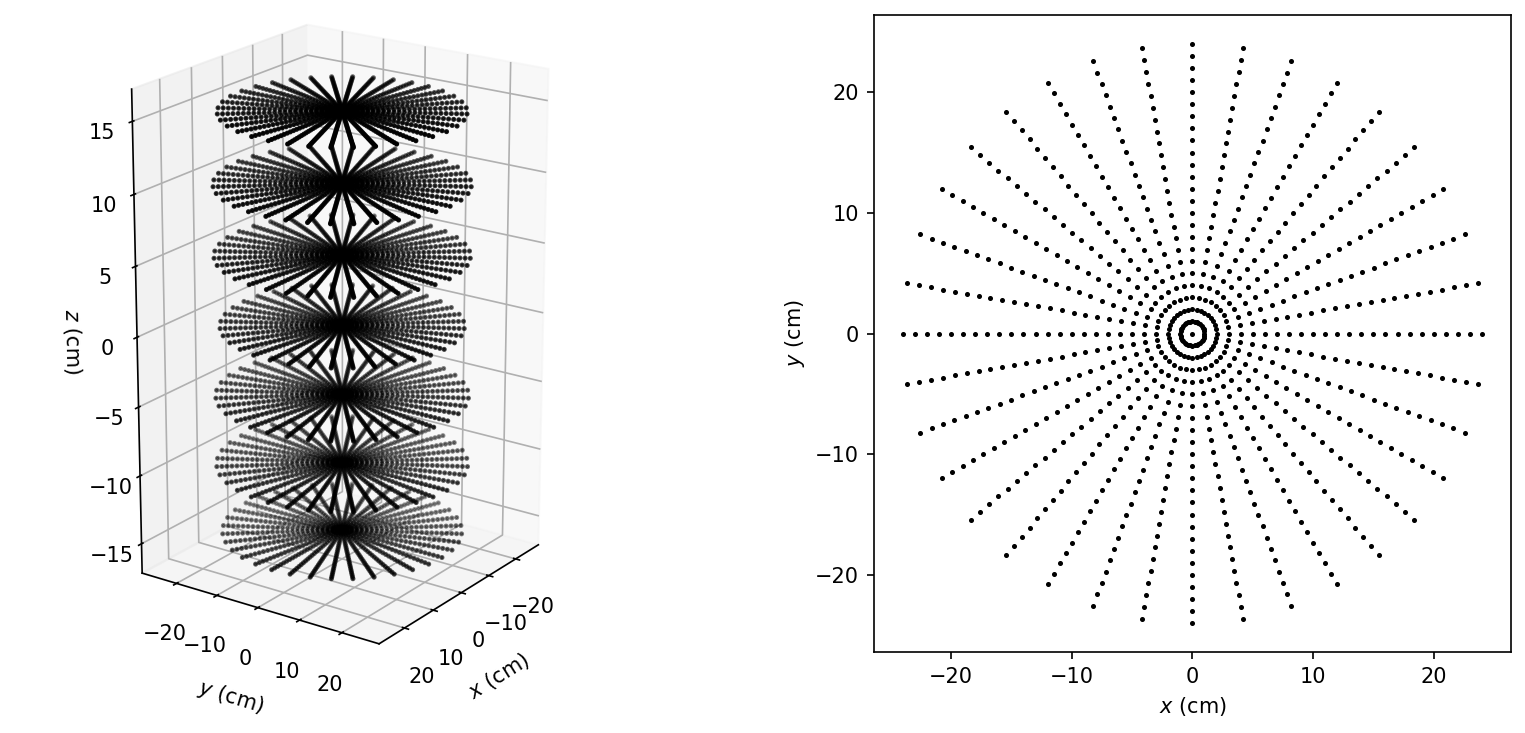}
    \caption{Distribution of the 6199 unique collocation points used during PINN training, comprising 25 radial rings, 7 axial planes, and 36 azimuthal positions per ring. The left panel shows the three-dimensional distribution, whereas the right panel shows the points in the $z=0$ plane.}
    \label{fig:collocation_geometry}
\end{figure}

In addition, collocation points were generated throughout the same cylindrical reconstruction volume. At these three-dimensional spatial locations, no magnetic field data were provided to the network; instead, the points were used to evaluate the spatial derivatives of the PINN-predicted magnetic field and enforce $\nabla\cdot\mathbf{B}=0$ and $\nabla\times\mathbf{B}=\mathbf{0}$ throughout the reconstruction volume. The collocation points were generated using radial, axial, and azimuthal spacings of $\Delta r=1~\mathrm{cm}$, $\Delta z=5~\mathrm{cm}$, and $\Delta\phi=10^{\circ}$, respectively. The radial coordinates extended from the central axis to $R=25~\mathrm{cm}$, and the axial coordinates were distributed over seven planes spanning $z=\pm15~\mathrm{cm}$. Each nonzero radial ring contained 36 uniformly spaced azimuthal positions. Duplicate coordinates and points coinciding with the mapper locations were removed, yielding 6199 unique collocation points,  as shown in Fig.~\ref{fig:collocation_geometry}.

Following training, the reconstructed field and its spatial derivatives were evaluated at the measurement locations. The nine first order derivatives form the magnetic field Jacobian,
\begingroup
\setlength{\abovedisplayskip}{5pt}
\setlength{\belowdisplayskip}{5pt}
\begin{equation}
\mathbf{J}_{B}
=
\begin{pmatrix}
\partial B_x/\partial x & \partial B_x/\partial y & \partial B_x/\partial z \\
\partial B_y/\partial x & \partial B_y/\partial y & \partial B_y/\partial z \\
\partial B_z/\partial x & \partial B_z/\partial y & \partial B_z/\partial z
\end{pmatrix}.
\label{eq:field_jacobian}
\end{equation}
\endgroup

The field reconstruction accuracy was quantified using the component-wise residuals. For each reconstruction method
$M\in\{\mathrm{PINN},\mathrm{SH}\}$ and magnetic field component
$r\in\{x,y,z\}$, the residual at the $i$th measurement location was defined as
\begin{equation}
\Delta B_{r,i}^{M}
=
B_{r,i}^{M}
-
B_{r,i}^{\mathrm{sim}},
\label{eq:field_residual}
\end{equation}
where $B_{r,i}^{M}$ is the field component reconstructed using method $M$, and
$B_{r,i}^{\mathrm{sim}}$ is the corresponding simulated reference value. The corresponding residual width was calculated as
\begin{equation}
\sigma_{ B_r^{M}}
=
\sqrt{
\frac{1}{N_m}
\sum_{i=1}^{N_m}
\left(
\Delta B_{r,i}^{M}
-
\overline{\Delta B_r^{M}}
\right)^2
},
\label{eq:residual_std}
\end{equation}
where $\overline{\Delta B_r^{M}}$ denotes the mean residual.

\subsubsection*{Measured STC field}

The STC consist of four independently powered coils labeled A, B, C, and D~\cite{Jared2023,Amara2023}. All four coils have an inner radius of $5.33~\mathrm{cm}$ and an outer radius of $6.61~\mathrm{cm}$. Coils A, B, C, and D have lengths of $10$, $15$, $25$, and $25~\mathrm{cm}$ and were operated at currents of $6.5$, $15$, $26$, and $100~\mathrm{mA}$, respectively. The operating currents can be scaled while maintaining the same relative ratios. The four coils are joined end-to-end, giving a total assembly length of $75~\mathrm{cm}$, to produce the required field.

\begin{figure}[htbp]
    \centering
    \includegraphics[width=\textwidth]
    {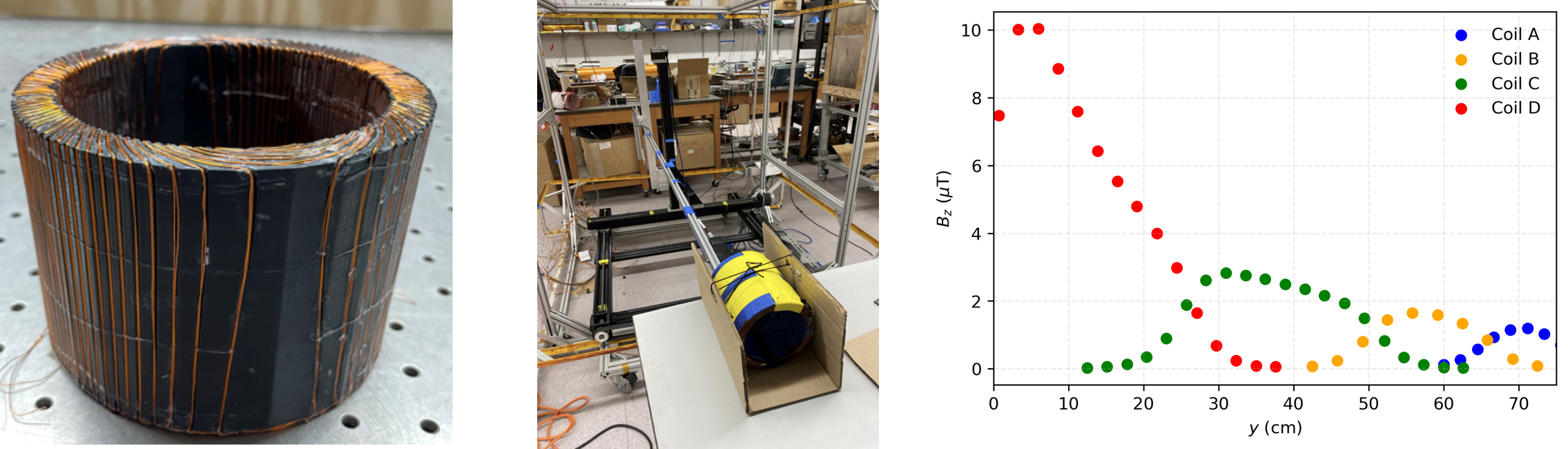}
    \caption{STC assembly, measurement setup, and measured magnetic-field profiles. The left panel shows coil A assembled from two 3D-printed halves, with grooves defining the copper-wire winding path. The center panel shows the experimental setup used at the University of Kentucky to measure the magnetic field produced by coil A. The right panel shows the measured axial-field component $B_z$ along the longitudinal coordinate $y$ at $x=z=0$ for coils A, B, C, and D.}
    \label{fig:stc_experiment}
\end{figure}

Figure~\ref{fig:stc_experiment} shows the assembled coil A and the experimental setup used for its magnetic field measurement. Coils B, C, and D were assembled and measured using the same procedure. The measurements were performed at the University of Kentucky between January and May 2025. The magnetic field data were collected using a Bartington three-axis fluxgate
magnetometer~\cite{BartingtonMag03} mounted on the movable arm of a stepper-motor controlled scanner,  following the mapping procedure described previously~\cite{Umit2022}.

The same PINN configuration, including the network architecture, loss formulation, and training hyperparameters , was employed for the reconstruction of the STC magnetic fields as was used for the reconstruction of the highly-uniform magnetic field, as a test of our PINN method's robustness across significantly different field profiles. A separate model was trained for each coil, with only the measured field data and collocation point locations changed to match the spatial range of the corresponding coil. For each model, the data loss was evaluated at the measurement locations, while the divergence and curl losses were evaluated at 2500 coil-specific collocation points. The collocation grid consisted of 100 uniformly spaced positions over the measured $y$ range and five positions along each of the $x$ and $z$ directions, spanning $\pm1.5~\mathrm{cm}$ about the measurement axis at $x=z=0$.

Each trained model was evaluated over the measured $y$ range of its corresponding coil to reconstruct the individual coil contribution. For comparison, the measured field of each coil was also interpolated over the same range using a cubic spline. 
The primary STC field component, $B_z$, was obtained by summing the four individual coil contributions: 

\begin{equation}
B_z^{\mathrm{total}}(y)
=
B_z^{A}(y)+B_z^{B}(y)+B_z^{C}(y)+B_z^{D}(y).
\label{eq:total_stc_field}
\end{equation}

\section*{Results}

\subsection*{Magnetic field reconstruction  of the highly-uniform magnetic field}

As described under Physics-Informed Neural Network Method, the PINN was trained by minimizing the data loss $\mathcal{L}_{\mathrm{data}}$ between the simulated and reconstructed magnetic field components at the mapper locations, as defined in Eq.~\eqref{eq:mse_loss}. The divergence and curl losses $\mathcal{L}_{\mathrm{div}}$ and $\mathcal{L}_{\mathrm{curl}}$, defined in Eqs.~\eqref{eq:div_loss} and~\eqref{eq:curl_loss}, respectively, were also computed at the collocation points to enforce $\nabla\cdot\mathbf{B}=0$ and $\nabla\times\mathbf{B}=\mathbf{0}$.

\begin{figure}[ht]
    \centering
    \includegraphics[width=\textwidth]{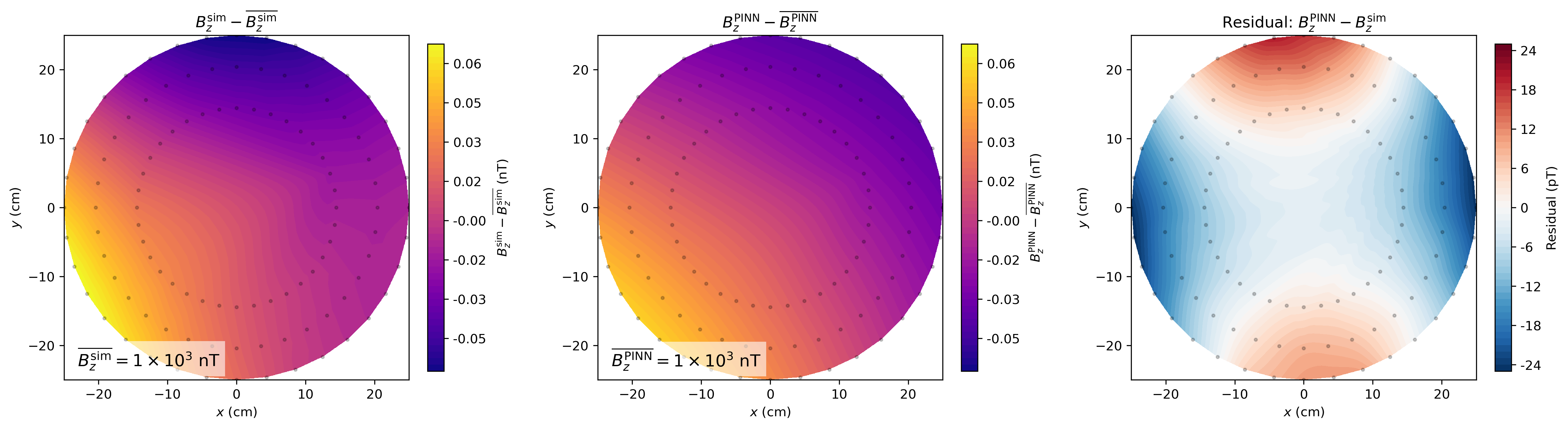}
    \caption{Comparison of the simulated $B_z$ field (left), the PINN-reconstructed field (center), and the residual $\Delta B_z=B_z^{\mathrm{PINN}}-B_z^{\mathrm{sim}}$ (right) at $z=0$. The simulated and reconstructed fields are reported in nT, whereas the residual is reported in pT. The points arranged on the three radial rings indicate the discrete mapper
locations.}
    \label{fig:pinn_heat_map_bz}
\end{figure}

The trained PINN was then used to reconstruct the three magnetic-field components, $B_x$, $B_y$, and $B_z$, at the 540 mapper locations defined by their $(x,y,z)$ coordinates. Figure~\ref{fig:pinn_heat_map_bz} compares the simulated and PINN-reconstructed axial field component $B_z$ in the central plane at $z=0$. Note that the observed asymmetry in the field uniformity is due to the added perturbative field, described previously. The residual shows a structured spatial variation, with the largest deviations occurring near the radial boundary and smaller deviations near the center of the mapping plane.

\begin{figure}[H]
    \centering
    \includegraphics[width=0.7\textwidth]{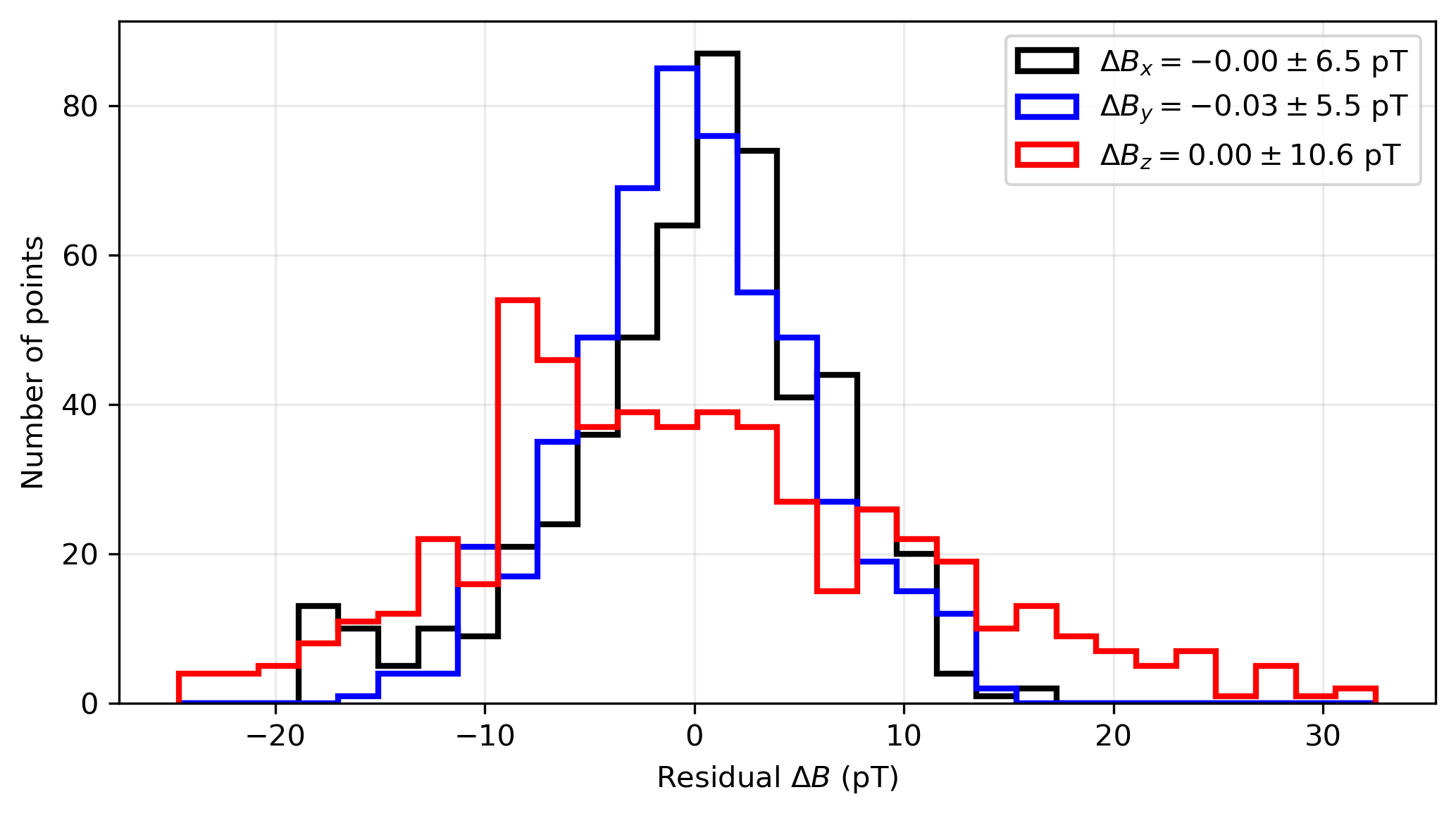}
    \caption{Distributions of the PINN reconstruction residuals, $\Delta B = B^{\mathrm{PINN}}-B^{\mathrm{sim}}$, for $B_x$, $B_y$, and $B_z$ at the 540 measurement locations. The legend gives the mean and standard deviation of each residual distribution in $\mathrm{pT}$.}

    \label{fig:three_residuals}
\end{figure}

Figure~\ref{fig:three_residuals} shows the residual distributions for the reconstructed field components $B_x$, $B_y$, and $B_z$. The residual widths, calculated using Eq.~\eqref{eq:residual_std}, are $6.5$, $5.5$, and $10.6~\mathrm{pT}$, respectively. Although $B_z$ has the largest absolute residual width, it is also the dominant field component because the coil is designed to produce a uniform field along the $z$ axis. The mean values of $B_x$, $B_y$, and $B_z$ are $-0.02$, $0.4$, and $1\times10^{3}~\mathrm{nT}$, respectively. The residual width of $B_z$ relative to its mean magnitude is therefore approximately $1.0\times10^{-5}$, substantially smaller than the corresponding ratios for $B_x$ and $B_y$.

\subsubsection*{Training convergence, reconstruction accuracy, and Maxwell consistency}

\begin{figure}[H]
    \centering
    \includegraphics[width=\textwidth]
    {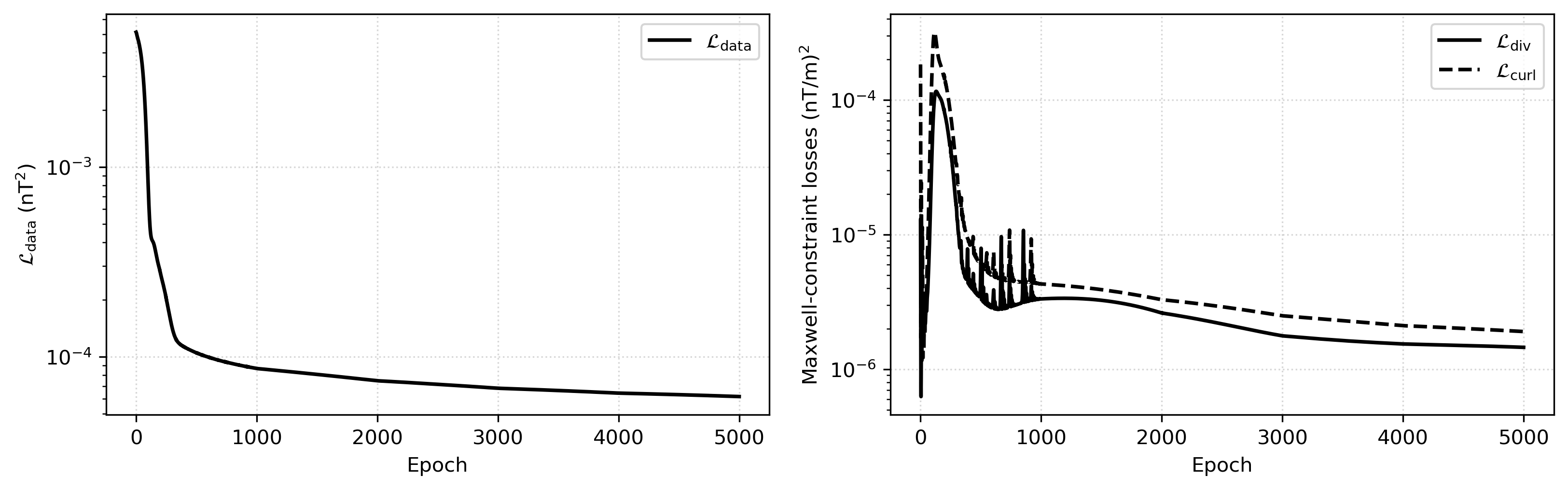}
    \caption{Convergence of the PINN loss terms. The left panel shows the mean-squared error between the simulated and PINN-reconstructed magnetic field components at the 540 mapper locations. The right panel shows the mean-squared divergence and curl losses evaluated at the 6199 collocation points.}
    \label{fig:loss_plot}
\end{figure}

Figure~\ref{fig:loss_plot} shows the evolution of the three loss terms during training: the data loss $\mathcal{L}_{\mathrm{data}}$ defined in Eq.~\eqref{eq:mse_loss}; the divergence loss $\mathcal{L}_{\mathrm{div}}$ defined in Eq.~\eqref{eq:div_loss}; and the curl loss $\mathcal{L}_{\mathrm{curl}}$ defined in Eq.~\eqref{eq:curl_loss}. The data loss $\mathcal{L}_{\mathrm{data}}$ decreased rapidly during the initial stage of training, falling by approximately two orders of magnitude within the first 500 epochs. It subsequently decreased more gradually and approached a stable value by the end of the 5000 training epochs. The divergence and curl losses initially increased and exhibited transient fluctuations while the network adjusted simultaneously to the field data and Maxwell constraints. After approximately 1000 epochs, both Maxwell-constraint losses decreased steadily.

At the end of training, $\sqrt{\mathcal{L}_{\mathrm{data}}}$ corresponded to a field RMSE of $7.9~\mathrm{pT}$ over the 540 mapper locations. Similarly, $\sqrt{\mathcal{L}_{\mathrm{div}}}$ and $\sqrt{\mathcal{L}_{\mathrm{curl}}}$ corresponded to RMS divergence and curl values of $1.2~\mathrm{pT\,m^{-1}}$ and $1.4~\mathrm{pT\,m^{-1}}$, respectively, over the 6199 collocation points.
 The simultaneous stabilization of the data and physics losses indicates that the trained model reached a balance between reconstructing the simulated field and satisfying the imposed Maxwell constraints.

\begin{table}[H]
\centering
\caption{Maxwell-constraint metrics for the PINN-reconstructed magnetic
field evaluated over the 540 mapper locations.}
\label{tab:physics_consistency}

\normalsize
\renewcommand{\arraystretch}{1.25}
\setlength{\tabcolsep}{0pt}

\begin{tabular*}{\textwidth}{
@{\extracolsep{\fill}}
ccccc
@{}
}
\toprule

\multicolumn{5}{c}{
Maxwell-constraint metrics
(mean $\pm$ standard deviation, $\mathrm{pT\,m^{-1}}$)} \\

\cmidrule(lr){1-5}

$\nabla\cdot\mathbf{B}$
& $(\nabla\times\mathbf{B})_x$
& $(\nabla\times\mathbf{B})_y$
& $(\nabla\times\mathbf{B})_z$
& $\lvert\nabla\times\mathbf{B}\rvert$ \\

\midrule

$-0.3 \pm 2.0$
& $\phantom{-}0.0 \pm 0.9$
& $\phantom{-}0.0 \pm 1.5$
& $\phantom{-}0.1 \pm 1.9$
& $\phantom{-}1.9 \pm 1.7$ \\

\bottomrule
\end{tabular*}
\end{table}

Following training, the reconstructed field and its spatial derivatives were evaluated at the 540 mapper locations. The magnetic field Jacobian defined in Eq.~\eqref{eq:field_jacobian} was used to calculate $\nabla\cdot\mathbf{B}$ and the three components of $\nabla\times\mathbf{B}$. These quantities provide a direct measure of the consistency of the reconstructed field with the source-free Maxwell constraints, $\nabla\cdot\mathbf{B}=0$ and $\nabla\times\mathbf{B}=\mathbf{0}$. Their spatial means and standard deviations are summarized in Table~\ref{tab:physics_consistency}. The mean divergence and curl components remain close to zero, with spatial variations at the $\mathrm{pT\,m^{-1}}$ level.

\subsubsection*{Comparison with SH reconstruction}

The spherical-harmonic (SH) expansion provides an analytical method for reconstructing the magnetic field and extracting its spatial gradients. For a direct comparison, the same simulated magnetic field values and spatial coordinates at the 540 mapper locations were used to train the PINN and fit the SH model. Using Eq.~\ref{eq:field_expansion}, the SH expansion coefficients were obtained by fitting the magnetic field data with modes up to $\ell_{\max}=3$.

\begin{figure}[H]
    \centering
    \includegraphics[width=\textwidth]{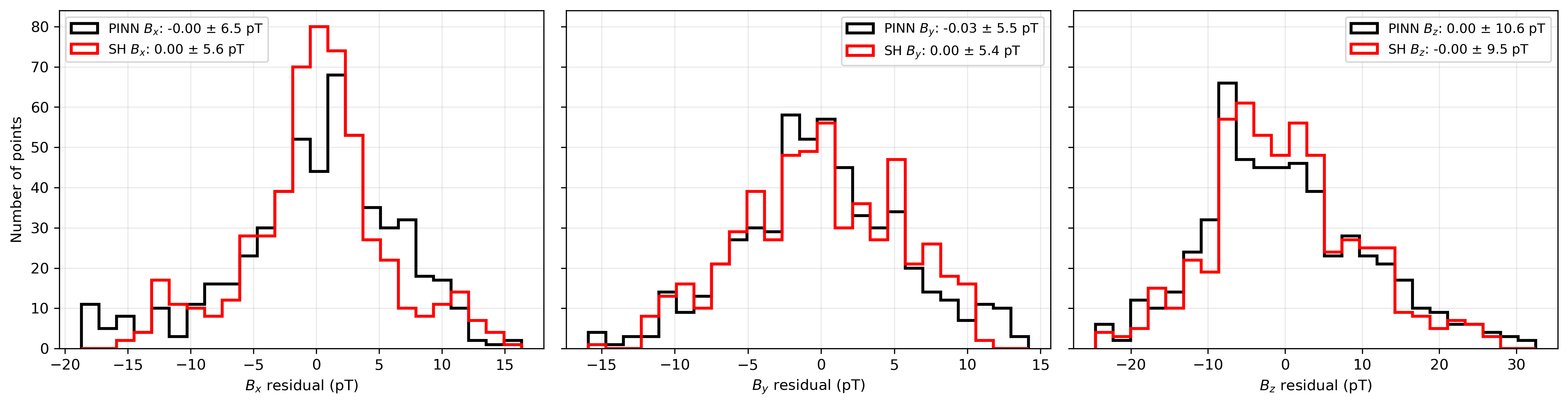}
    \caption{Residual distributions for the PINN and SH reconstructions of $B_x$, $B_y$, and $B_z$ evaluated at the 540 mapper locations. The residuals represent the differences between the reconstructed and simulated field values at these locations. The legends give the mean and standard deviation of each distribution in $\mathrm{pT}$.}
    \label{fig:pinn_sh_residuals}
\end{figure}

Figure~\ref{fig:pinn_sh_residuals} compares the component-wise residual distributions obtained using the PINN and SH reconstructions. Both methods reproduce the simulated field with residual distributions centered close to zero. The PINN residual widths were $6.5$, $5.5$, and $10.6~\mathrm{pT}$ for $B_x$, $B_y$, and $B_z$, respectively, compared with $5.6$, $5.4$, and $9.5~\mathrm{pT}$ for the SH reconstruction. Thus, the SH fit produced slightly narrower residual distributions, although both methods provided comparable reconstruction accuracy.

\begin{figure}[H]
    \centering
    \includegraphics[width=\textwidth]
    {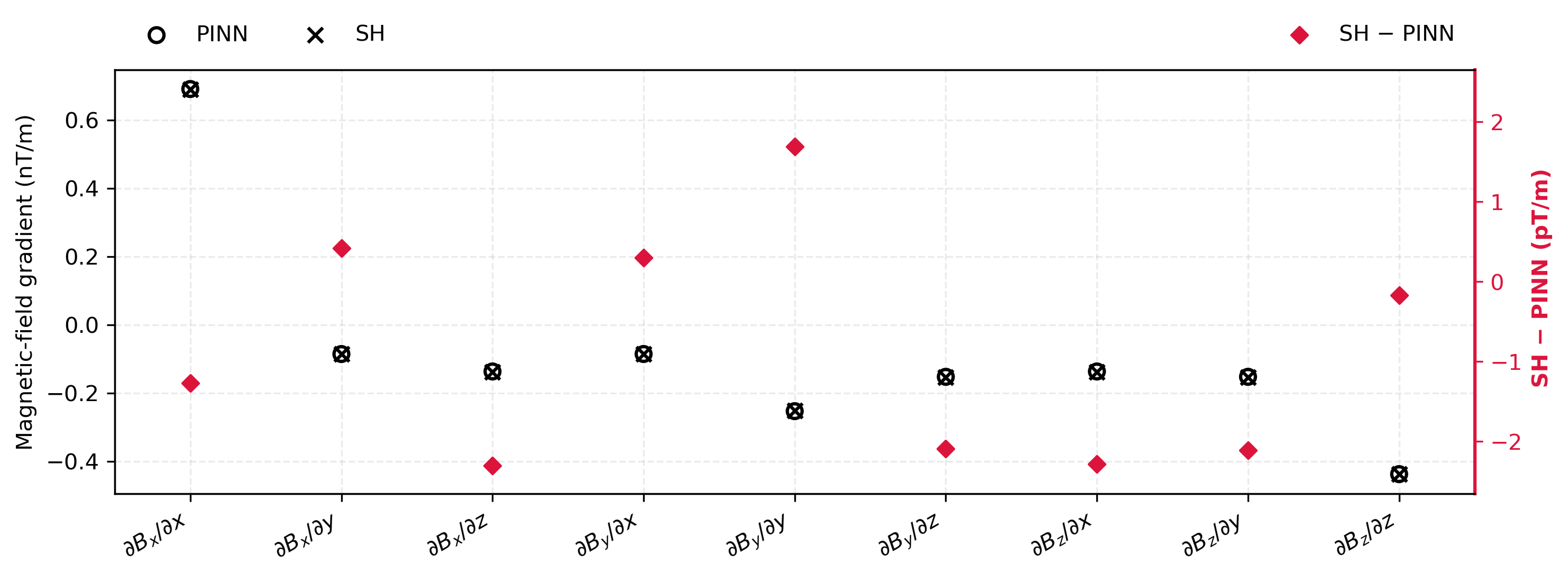}
    \caption{Comparison of the nine first order magnetic field gradient components obtained using the PINN and SH methods. The black markers show the spatially averaged gradient components in $\mathrm{nT\,m^{-1}}$ on the left vertical axis, while the red diamonds show the corresponding differences, $\mathrm{PINN}-\mathrm{SH}$, in $\mathrm{pT\,m^{-1}}$ on the right vertical axis.}
    \label{fig:pinn_sh_gradients}
\end{figure}

Following reconstruction, the nine first order spatial derivatives of the PINN field were evaluated at the same 540 locations. For each derivative term, the values at the 540 locations were averaged separately, yielding the spatial means of the nine PINN gradient components. The corresponding SH gradients were obtained analytically from the fitted expansion coefficients, as described in the section Spherical-harmonic (SH) reconstruction. The PINN and SH results for all first order magnetic field gradient components, shown in Fig.~\ref{fig:pinn_sh_gradients}, agreed, with differences of only a few $\mathrm{pT\,m^{-1}}$. The largest absolute difference was approximately $2.3~\mathrm{pT\,m^{-1}}$, which was small relative to the magnitude of the corresponding gradient. Furthermore, the corresponding cross-derivative pairs obtained using the PINN and SH methods, $\partial B_x/\partial y$ and $\partial B_y/\partial x$, $\partial B_x/\partial z$ and $\partial B_z/\partial x$, and $\partial B_y/\partial z$ and $\partial B_z/\partial y$, agreed. This  agreement demonstrates that the PINN reproduces not only the magnetic field values but also the first order spatial derivatives obtained using the established SH reconstruction.

\subsubsection*{Effect of the number of measurement points}

The number of field measurements is an important practical consideration in an experimental mapping campaign, impacting the overall measurement time, the time between measurements in different physical regions, etc. We therefore investigated the sensitivity of the PINN reconstruction to the number of measurement points. In addition to the standard configuration of 540 measurement points, models were trained using 135 and 1080 measurement points. The radial and axial coordinates were unchanged, while the azimuthal spacings were $40^\circ$ for the 135 point dataset and $5^\circ$ for the 1080 point dataset. The magnetic fields at these locations were obtained using the same approach as for the 540 measurement point dataset. These configurations were used to study how doubling the number of measurement points or reducing it by a factor of four affects the field reconstruction accuracy and the extracted gradients. The network architecture, activation function, collocation points, loss weights, optimizer settings, and number of training epochs were held fixed so that only the number of measurement points was varied.

\begin{figure}[H]
    \centering
    \includegraphics[width=\textwidth]
    {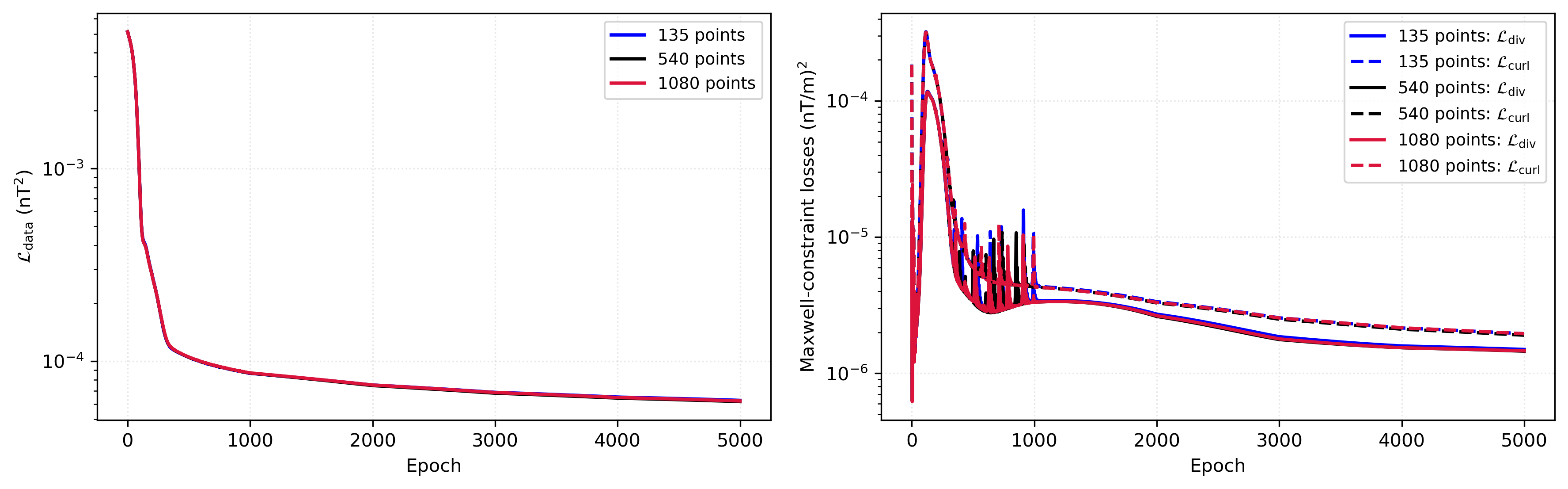}
    \caption{PINN loss convergence for 135, 540, and 1080 measurement points. The left panel shows the data loss, while the right panel shows the divergence (solid) and curl (dashed) losses evaluated at the 6199 collocation points.}
    \label{fig:mapper_loss_comparison}
\end{figure}

Figure~\ref{fig:mapper_loss_comparison} compares the loss histories for the three number of  measurement point configurations. All three models exhibited similar convergence behavior in both the magnetic-field data loss and the Maxwell-constraint losses. Thus, over the range considered, changing the number of measurement points did not substantially affect the stability or overall convergence of the training.

The trained models were used to reconstruct the field using each number of measurement points. The physical consistency and reconstruction accuracy evaluated at 135, 540, and 1080 measurement locations are summarized in Table~\ref{tab:mapper_point_comparison}. For all three measurement point configurations, the mean divergence and individual curl components remained close to zero, while their spatial standard deviations were comparable. However,  the mean magnitude  of the curl was approximately $1.9~\mathrm{pT\,m^{-1}}$ for all three configurations of measurement points.

\begin{table}[H]
\centering
\caption{Physical consistency and reconstruction accuracy for different
numbers of measurement points. The Maxwell-constraint quantities are
reported as spatial mean $\pm$ standard deviation in
$\mathrm{pT\,m^{-1}}$. The field residuals are reported as standard
deviations in $\mathrm{pT}$.}
\label{tab:mapper_point_comparison}

\small
\renewcommand{\arraystretch}{1.3}
\setlength{\tabcolsep}{3pt}

\resizebox{\textwidth}{!}{%
\begin{tabular}{
C{1.7cm}
*{5}{C{1.65cm}}
*{3}{C{1.25cm}}
}
\toprule

& \multicolumn{5}{c}{%
Maxwell-constraint metrics
(mean $\pm$ standard deviation, $\mathrm{pT\,m^{-1}}$)}
& \multicolumn{3}{c}{%
Field residuals ($\mathrm{pT}$)} \\

\cmidrule(lr){2-6}
\cmidrule(lr){7-9}

Measurement points
& $\nabla\cdot\mathbf{B}$
& $(\nabla\times\mathbf{B})_x$
& $(\nabla\times\mathbf{B})_y$
& $(\nabla\times\mathbf{B})_z$
& $\lvert\nabla\times\mathbf{B}\rvert$
& $\sigma_{B_x}$
& $\sigma_{B_y}$
& $\sigma_{ B_z}$ \\

\midrule

135
& $-0.3 \pm 1.9$
& $\phantom{-}0.0 \pm 0.8$
& $\phantom{-}0.0 \pm 1.4$
& $\phantom{-}0.1 \pm 1.8$
& $\phantom{-}1.9 \pm 1.6$
& $6.6$
& $5.6$
& $10.7$ \\

540
& $-0.3 \pm 2.0$
& $\phantom{-}0.0 \pm 0.9$
& $\phantom{-}0.0 \pm 1.5$
& $\phantom{-}0.1 \pm 1.9$
& $\phantom{-}1.9 \pm 1.7$
& $6.5$
& $5.5$
& $10.6$ \\

1080
& $-0.3 \pm 2.0$
& $\phantom{-}0.0 \pm 0.9$
& $\phantom{-}0.0 \pm 1.5$
& $\phantom{-}0.1 \pm 1.9$
& $\phantom{-}1.9 \pm 1.7$
& $6.6$
& $5.6$
& $10.6$ \\

\bottomrule
\end{tabular}%
}
\end{table}

The component-wise residual widths also showed little variation with the number of measurement points. Specifically, $\sigma_{ B_x}$ ranged from $6.5$ to $6.6~\mathrm{pT}$, $\sigma_{B_y}$ ranged from $5.5$ to $5.6~\mathrm{pT}$, and $\sigma_{ B_z}$ ranged from $10.6$ to $10.7~\mathrm{pT}$. Thus, reducing the number of measurement points from 540 to 135 did not degrade the reconstruction for the simulated field and sampling geometry considered. Similarly, increasing the number from 540 to 1080 produced no substantial improvement under the same training conditions.

The stability of the extracted first-order gradients was evaluated using 135, 540, and 1080 measurement locations. 
For the PINN reconstruction, all nine gradient components remained nearly unchanged as the number of measurement points was varied, with a maximum variation of approximately $0.2~\mathrm{pT\,m^{-1}}$. The corresponding PINN cross-derivative pairs, $\partial B_x/\partial y$ and $\partial B_y/\partial x$, $\partial B_x/\partial z$ and $\partial B_z/\partial x$, and $\partial B_y/\partial z$ and $\partial B_z/\partial y$, also agreed. Similarly, the SH-derived gradients exhibited a maximum variation of approximately $0.2~\mathrm{pT\,m^{-1}}$ and preserved the expected cross-derivative symmetries. These results indicate that the extracted first order gradients are insensitive to whether 135, 540, or 1080 measurement points are used.

\subsection*{STC magnetic field reconstruction}

\begin{figure}[H]
    \centering
    \includegraphics[width=0.98\linewidth]
    {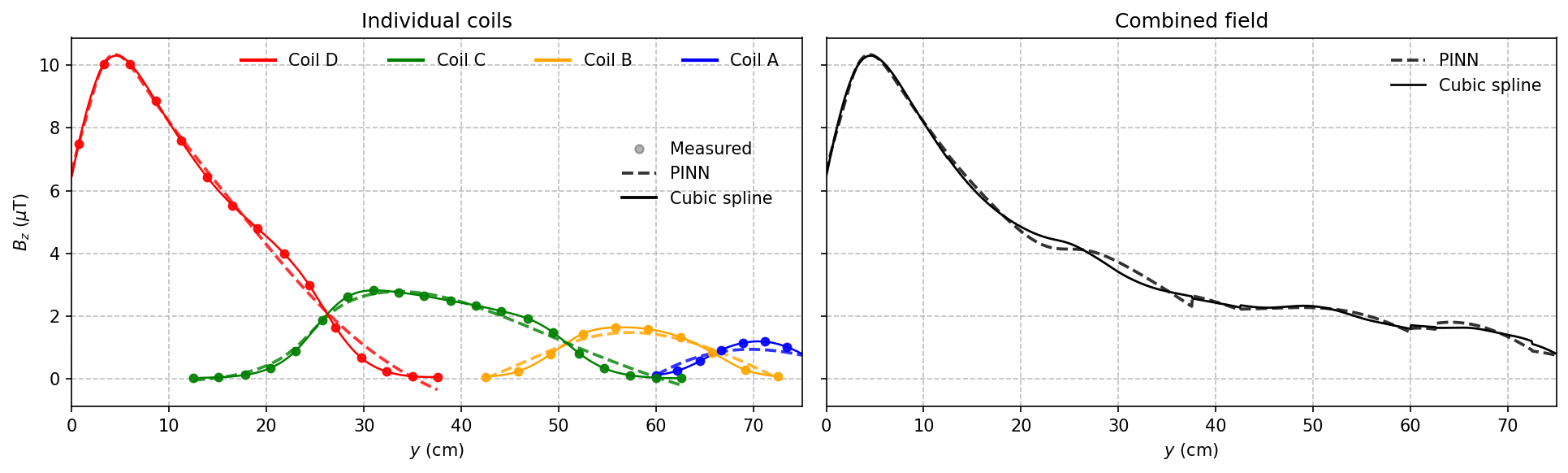}
    \caption{Reconstruction of the STC axial-field component $B_z$ along $y$ at $x=z=0$. The left panel shows the measured values and individual PINN and cubic-spline reconstructions for coils A, B, C, and D. The right panel shows the total field obtained by summing the reconstructed contributions from all four coils.}
    \label{fig:stc_reconstruction}
\end{figure}

Figure~\ref{fig:stc_reconstruction} shows that the PINN reconstructions follow the measured profiles and agree closely with the cubic-spline interpolations for all four coils. Coil D produces the largest field, reaching approximately $10.3~\mu\mathrm{T}$ near $y=5~\mathrm{cm}$, while the maximum fields produced by coils C, B, and A are approximately $2.8$, $1.6$, and $1.2~\mu\mathrm{T}$, respectively.

The combined field reaches approximately $10.3~\mu\mathrm{T}$ near the beginning of the transport path and subsequently decreases to approximately $1~\mu\mathrm{T}$ near the end of the $75~\mathrm{cm}$ assembly. The PINN and cubic-spline profiles exhibit the same overall behavior, with small differences primarily in the region of  overlap between adjacent coil contributions. These results indicate that the independently trained PINN models reproduce both the individual coil fields and the combined STC field consistently with cubic-spline interpolation.

\section*{Conclusion}
We have developed a physics-informed neural network which reconstructs magnetic field components and gradients from discrete measurement points subject to Maxwell’s equations. We have demonstrated the utility of our method through application to two specific examples of significantly different field profiles.  Our results have been benchmarked by a  comparison of our results with a reconstruction based on spherical harmonics. Our method may be of broad interest to precision measurement experiments requiring detailed knowledge of magnetic fields.

\section*{Data Availability}
The datasets used in this study are available in the GitHub repository at \url{https://github.com/pcad222/PINN-Magnetic-Field-Reconstruction}.

\section*{Acknowledgements}

This work was supported in part by the U.S. Department of Energy under Award No.~DE-SC0014622, and also by the Office of the Vice President for Research at the University of Kentucky through support for one of us (F.B.H.).

\section*{Author contributions}

P.A. and A.T.-B. developed the concept. P.A. performed the PINN calculations. R.B., J.B., D.B., R.M., and P.A.P. developed the magnetic-field models in COMSOL and collected the data for the two specific examples. F.B.H., M.F.H., A.R., and S.S. reviewed and assessed the performance of the models. B.P. supervised the research. All authors interpreted the results, contributed to the writing of the manuscript, and reviewed and approved the final version.

\section*{Competing interests}

The authors declare no competing interests.



\bibliographystyle{naturemag}
\bibliography{references}

@article{Umit2022,
  author  = {Coskun, Umit H. and Sel, Bilgehan and Plaster, Brad},
  title   = {Magnetic field mapping of inaccessible regions using physics-informed neural networks},
  journal = {Sci. Rep.},
  volume  = {12},
  pages   = {12858},
  year    = {2022},
  doi     = {10.1038/s41598-022-15777-4},
  url     = {https://doi.org/10.1038/s41598-022-15777-4}
}

@article{Chupp2019,
  title     = {Electric dipole moments of atoms, molecules, nuclei, and particles},
  author    = {Chupp, T. E. and Fierlinger, P. and Ramsey-Musolf, M. J. and Singh, J. T.},
  journal   = {Reviews of Modern Physics},
  volume    = {91},
  number    = {1},
  pages     = {015001},
  year      = {2019},
  publisher = {American Physical Society},
  doi       = {10.1103/RevModPhys.91.015001},
  url       = {https://doi.org/10.1103/RevModPhys.91.015001}
}

@article{Pendlebury2004,
  author  = {Pendlebury, J. M. and others},
  title   = {Geometric-phase-induced false electric dipole moment signals for particles in traps},
  journal = {Phys. Rev. A},
  volume  = {70},
  pages   = {032102},
  year    = {2004},
  doi     = {10.1103/PhysRevA.70.032102},
  url     = {https://doi.org/10.1103/PhysRevA.70.032102}
}

@article{Lamoreaux2005,
  author  = {Lamoreaux, S. K. and Golub, R.},
  title   = {Detailed discussion of a linear electric field frequency shift induced in confined gases by a magnetic field gradient: Implications for neutron electric-dipole-moment experiments},
  journal = {Phys. Rev. A},
  volume  = {71},
  pages   = {032104},
  year    = {2005},
  doi     = {10.1103/PhysRevA.71.032104},
  url     = {https://doi.org/10.1103/PhysRevA.71.032104}
}

@article{Barabanov2006,
  author  = {Barabanov, A. L. and Golub, R. and Lamoreaux, S. K.},
  title   = {Electric dipole moment searches: Effect of linear electric field frequency shifts induced in confined gases},
  journal = {Phys. Rev. A},
  volume  = {74},
  pages   = {052115},
  year    = {2006},
  doi     = {10.1103/PhysRevA.74.052115},
  url     = {https://doi.org/10.1103/PhysRevA.74.052115}
}

@article{Nouri2014,
  author  = {Nouri, N. and Plaster, B.},
  title   = {Systematic optimization of exterior measurement locations for the determination of interior magnetic field vector components in inaccessible regions},
  journal = {Nucl. Instrum. Methods Phys. Res. A},
  volume  = {767},
  pages   = {92--98},
  year    = {2014},
  doi     = {10.1016/j.nima.2014.08.026},
  url     = {https://doi.org/10.1016/j.nima.2014.08.026}
}

@article{Brad2014,
  author  = {Nouri, N. and Plaster, B.},
  title   = {Sensitivity requirements for accessing interior magnetic field vector components in neutron electric dipole moment experiments via exterior boundary-value measurements},
  journal = {J. Instrum.},
  volume  = {9},
  pages   = {P11009},
  year    = {2014},
  doi     = {10.1088/1748-0221/9/11/P11009},
  url     = {https://doi.org/10.1088/1748-0221/9/11/P11009}
}

@article{Nouri2015,
  author  = {Nouri, N. and others},
  title   = {A prototype vector magnetic field monitoring system for a neutron electric dipole moment experiment},
  journal = {J. Instrum.},
  volume  = {10},
  number  = {12},
  pages   = {P12003},
  year    = {2015},
  doi     = {10.1088/1748-0221/10/12/P12003},
  url     = {https://doi.org/10.1088/1748-0221/10/12/P12003}
}

@article{Abel2019,
  author  = {Abel, C. and others},
  title   = {Magnetic-field uniformity in neutron electric-dipole-moment experiments},
  journal = {Phys. Rev. A},
  volume  = {99},
  pages   = {042112},
  year    = {2019},
  doi     = {10.1103/PhysRevA.99.042112},
  url     = {https://doi.org/10.1103/PhysRevA.99.042112}
}

@article{Abel2022,
  author  = {Abel, C. and others},
  title   = {Mapping of the magnetic field to correct systematic effects in a neutron electric dipole moment experiment},
  journal = {Phys. Rev. A},
  volume  = {106},
  pages   = {032808},
  year    = {2022},
  doi     = {10.1103/PhysRevA.106.032808},
  url     = {https://doi.org/10.1103/PhysRevA.106.032808}
}

@article{Raissi2019,
  author  = {Raissi, Maziar and Perdikaris, Paris and Karniadakis, George Em},
  title   = {Physics-informed neural networks: A deep learning framework for solving forward and inverse problems involving nonlinear partial differential equations},
  journal = {J. Comput. Phys.},
  volume  = {378},
  pages   = {686--707},
  year    = {2019},
  doi     = {10.1016/j.jcp.2018.10.045},
  url     = {https://doi.org/10.1016/j.jcp.2018.10.045}
}

@article{baydin2018automatic_diff,
  author  = {Baydin, Atilim Gunes and Pearlmutter, Barak A. and Radul, Alexey Andreyevich and Siskind, Jeffrey Mark},
  title   = {Automatic differentiation in machine learning: A survey},
  journal = {J. Mach. Learn. Res.},
  volume  = {18},
  number  = {153},
  pages   = {1--43},
  year    = {2018},
  url     = {https://www.jmlr.org/papers/v18/17-468.html}
}

@inproceedings{pytorch2019,
  author    = {Paszke, Adam and others},
  title     = {{PyTorch}: An imperative style, high-performance deep learning library},
  booktitle = {Advances in Neural Information Processing Systems},
  volume    = {32},
  year      = {2019},
  url       = {https://proceedings.neurips.cc/paper/2019/hash/bdbca288fee7f92f2bfa9f7012727740-Abstract.html}
}

@phdthesis{Ryan2018,
  author = {Dadisman, James Ryan},
  title  = {Magnetic field design to reduce systematic effects in neutron electric dipole moment measurements},
  school = {University of Kentucky},
  type   = {Doctoral dissertation},
  year   = {2018},
  doi    = {10.13023/ETD.2018.094},
  url    = {https://doi.org/10.13023/ETD.2018.094}
}

@phdthesis{Jared2023,
  author = {Brewington, Jared},
  title  = {Design of the high uniform magnetic field and spin-transport magnetic field coils for the {Los Alamos National Laboratory} neutron electric dipole moment experiment},
  school = {University of Kentucky},
  type   = {Doctoral dissertation},
  year   = {2023},
  doi    = {10.13023/etd.2023.226},
  url    = {https://doi.org/10.13023/etd.2023.226}
}

@phdthesis{Amara2023,
  author = {Palamure, Piya Amara},
  title  = {The design of primary holding magnet for the {LANL} neutron electric dipole moment experiment},
  school = {University of Kentucky},
  type   = {Doctoral dissertation},
  year   = {2023},
  doi    = {10.13023/etd.2023.225},
  url    = {https://doi.org/10.13023/etd.2023.225}
}

@article{Crawford2021,
  author  = {Crawford, Christopher B. and Straley, Joseph P.},
  title   = {Improved prescription for winding an electromagnet},
  journal = {Rev. Sci. Instrum.},
  volume  = {92},
  number  = {12},
  pages   = {124707},
  year    = {2021},
  doi     = {10.1063/5.0063057},
  url     = {https://doi.org/10.1063/5.0063057}
}

@article{Crawford2021_design,
  author  = {Crawford, C. B.},
  title   = {The physical meaning of the magnetic scalar potential and its use in the design of hermetic electromagnetic coils},
  journal = {Rev. Sci. Instrum.},
  volume  = {92},
  number  = {12},
  pages   = {124703},
  year    = {2021},
  doi     = {10.1063/5.0063054},
  url     = {https://doi.org/10.1063/5.0063054}
}

@unpublished{STC_inpreparation,
  author = {Bhattarai, R. and others},
  title  = {Manuscript in preparation},
  year   = {2026}
}

@misc{COMSOL61,
  author       = {{COMSOL AB}},
  title        = {{COMSOL Multiphysics} v. 6.1},
  howpublished = {\url{https://www.comsol.com}},
  address      = {Stockholm, Sweden}
}

@misc{BartingtonMag03,
  author = {{Bartington Instruments Ltd.}},
  title  = {{Mag-03} three-axis fluxgate magnetometer},
  url    = {https://www.bartington.com/products/mag-03/}
}

\end{document}